\documentstyle{l-aa}

\newif\ifnofig\nofigtrue   
\newif\ifthisfig
\ifnofig\else\input{psfig}\psfigurepath{user2:[lecavelier.photo]}\fi

\def\setfigdefault{\ifnofig\thisfigfalse\else\thisfigtrue\fi}
\setfigdefault

\newcommand{\bp}{$\beta\:$Pictoris}

\begin{document}
\thesaurus{08 (08.09.2 $\beta\:$Pic; 08.03.4; 08.16.2; 03.13.2; 08.22.3)}
\title{$\beta$~Pictoris : Evidence of light variations.
\thanks{Based on observations obtained at the European Southern
Observatory (ESO), La Silla (Chile)}}

\author{
A. Lecavelier des Etangs \inst{1} \and
M. Deleuil \inst{2} \and
A. Vidal-Madjar \inst{1} \and
R. Ferlet \inst{1} \and
C. Nitschelm \inst{3} \and
B. Nicolet \inst{3} \and
A.M. Lagrange-Henri \inst{4}
}

\offprints{A. Lecavelier des Etangs}
\institute{
Institut d'Astrophysique de Paris, 98 Boulevard Arago,
F-75014 Paris, France
\and
Laboratoire d'Astronomie Spatiale du CNRS, BP 8, F-13376 Marseille
Cedex 12, France
\and
Observatoire de Gen$\rm{\grave e}$ve,
CH-1290 Sauverny, Suisse
\and
Groupe d'Astrophysique de Grenoble, CERMO BP53X, F-38041 Grenoble Cedex, France
}

\date{Received; accepted}

\maketitle

\begin{abstract}

We have analyzed \bp\ photometric measurements obtained from La Silla
by the Geneva Observatory from 1975 to 1992.
These data show evidence of
variations in the brightness of the star, with no color dependency. Here, we
demonstrate that the light variations are present on long as well as on short
time scales.

On a long time scale, we show that the apparent magnitude of \bp\ decreased by
0$^{\scriptscriptstyle m}\!\!.$011 $\pm$ 0$^{\scriptscriptstyle m}\!\!.$004
from 1979 to 1982. Moreover, when we consider all the
measurements, the chance that there is no variation at all can be estimated to
be less than $10^{-4}$. On short time scales there is a peculiar
feature observed during about 30 days; the variations may be as high as
0$^{\scriptscriptstyle m}\!\!.$04
magnitude. A maximum entropy reconstruction of the photometric data is
tentatively proposed and some physical interpretations are presented.


\end{abstract}

\section{Introduction}

Following the Geneva Observatory photometry measurements obtained from La
Silla (Rufener 1989), we were able to analyze the \bp\ (HR~2020, HD~39060)
photometric variations. The data presented here were obtained from Nov 18, 1975
to Feb 27, 1992 (Julian Day 2442734 to 2448679)\footnote{data-file available
electronically at IAP via anonymous ftp on corton.iap.fr
/pub/from\_users/lecaveli/}.

In 1975, \bp\ was used by the Geneva Observatory as a reference star. But,
although the "Vega-like" phenomenon was not yet known (Aumann et al. 1984), it
was noted that \bp\ showed some variations, and it was eliminated from the
list of reference stars in 1982. Unfortunately, the consequence was that this
star was not observed any more until 1988, when we suggested to start again its
photometric survey. Indeed, \bp\ was noted as a very peculiar star in 1984
(Smith \& Terrile 1984). In ten years a large number of observations of the
disk were performed in optical imagery with coronographs (Smith \& Terrile
1987, Paresce \& Burrows 1987, Golimowsky et al. 1993) or anti-blooming CCD
(Lecavelier des Etangs et al. 1993), in infrared photometry (Backman et al.
1992) and imagery (Lagage \& Pantin 1994). Meanwhile spectroscopic studies
carried out on a large range of wavelengths have led to analysis of the gaseous
counterpart of the disk: in the UV with IUE (Kondo \& Bruhweiler 1985,
Lagrange-Henri et al. 1989, Deleuil et al. 1993) and HST (Boggess et al. 1991,
Vidal-Madjar et al. 1994), in optical (Hobbs et al. 1985, Vidal-Madjar et al.
1986, Ferlet et al. 1993, Crawford et al. 1994) and infrared (Knacke et al.
1993, Aitken et al. 1993). The circumstellar spectral lines show strong
evidence of variations, which are presently interpreted as comets falling onto
the star (Beust et al. 1991, Beust \& Tagger 1993). The last HST observations
(Vidal-Madjar et al. 1994) strongly confirm our model and its predictions. The
presence of numerous large bodies in the \bp\ circumstellar disk is a more and
more reasonable hypothesis.

However, analysis of photometric variation is a new approach which has never
been tried for \bp. As already mentioned, we proposed in 1989 to return \bp\ to
the list of stars observed by the Geneva Observatory. Despite the lack of
photometric measurements from 1983 to 1988, the data collected until now cover
a very long time base that amounts to almost twenty years. It is now possible
to extract some important and new information, and especially statistical
information derived from data taken over a large time scale .

The data are presented in Sect.~\ref{presentation}. In Sect.~\ref{long term}
and \ref{short term}, we shall show the analysis of the variations in terms of
long and short time scales. Some physical interpretations are presented in
Sect.~\ref{interpretations}

\section{Presentation of the data}
\label{presentation}

Throughout the paper, we shall subtract a constant (2440000) from the
Julian Day,
for simplicity, considering the four last significant digits; for example
Julian Day 2444918 will be designated as JD 4918. The measurements are composed
of one V magnitude and six color index measurements corrected for atmospheric
extinction: V, U-B, V-B, B$_1$-B, B$_2$-B, V$_1$-B, G-B (Rufener \& Nicolet
1988). Thus, the same tests can be performed on all these seven quantities.
But, as we shall see, except in the section~\ref{from 7000}, the color
determinations do not show variations.

The magnitude measurements are associated with a quality factor from 0 to 4 in
which 4 is the best quality measurement. If we consider all the data
irrespective of the quality factor, the results are the same as those presented
here, but the probabilities are not so well distributed (see below), so the
uncertainties cannot be estimated easily. Among the 238 measurements of \bp, we
considered only those 155 measurements that have a quality greater than or
equal to~3.

All the statistical tests we performed on these data were also performed on 8
other stars to check the reality of the variations observed and to confirm that
the probabilities are well distributed. These stars are the following with the
number of measurements with quality greater than or equal to~3 in brackets:
HR 10 (64), HR 33 (157), HR 1801 (82), HR 2154 (198), HR 2174 (41),
HR 6389 (67), HR 6519 (6), HR 7316 (216).

\section{Long term variations}
\label{long term}

\subsection{Estimate of the parameters and the significance levels}

In this section, we consider each V measurements ($V_i$) as a function of the
time ($t_i$). In order to estimate the variations of the magnitude measurements
and the significance of these variations, we used least square methods, and
tried to fit the data with two models, that is first and second
degree polynomials. In
each case, we compute the variance of the parameter estimates, so we are able
to conclude if the variations are significant or not. In other words, we
estimate the probability that the result is only an artifact of statistical
random variations, which would mean that the magnitude data do not reveal real
variations, and represent only randomly distributed deviations around the real
magnitude of the star.

The two models are:
$
V_1  =  a_1  +  b_1\cdot t
$\ and
$
V_2  =  a_2  +  b_2\cdot t + c_2\cdot t^2
$.
In the first model, $s^2_{b_1}$ is an estimate of the variance of $b_1$:
$
s^2_{b_1} = Q_{1}/(N-2)\sum_{i}^{N} ( t_i-\overline t )^{2}
$
with $Q_1=\sum_{i}^{N} \left( V_i-V_1(t_i)\right)^{2}$.

The probability that the estimator $|\hat b_1|$ is as high as the calculated
estimate $|b_1(V_i,t_i)|$, under the hypothesis that $b_1$ is zero, gives a
good estimation of the probability that there is no signal. We use the fact
that $\hat b_{1}/s_{b_{1}}$ follows a Student law with N-2 degrees of freedom,
if there are N measurements.

In order to check if the addition of the $t^2$ term in the second model
improves the goodness of fit of the first model, we evaluate
$F=(Q_1-Q_2)(N-3)/Q_2 $ where $Q_2=\sum_{i}^{N}(V_i-V_2(t_i))^{2}$. If the
$t^2$ term does not improve the fit, then $Q_1=Q_2$ and $F=0$. On the other
hand, if $F$ is very high, thus $Q_1\gg Q_2$, the data will strongly depend on
the $t^2$ term. $\hat F$ follows a Fisher-Snedecor law with 1 and (N-3) degrees
of freedom. The probability that $\hat F$ is as high as the calculated one is
an estimate of the probability that the magnitude of \bp\ does not depend on
the $t^2$ term.

\subsection{Data until 1982}
\label{from 3900}

\begin{figure}[tbp]
\ifthisfig
\psfig{file=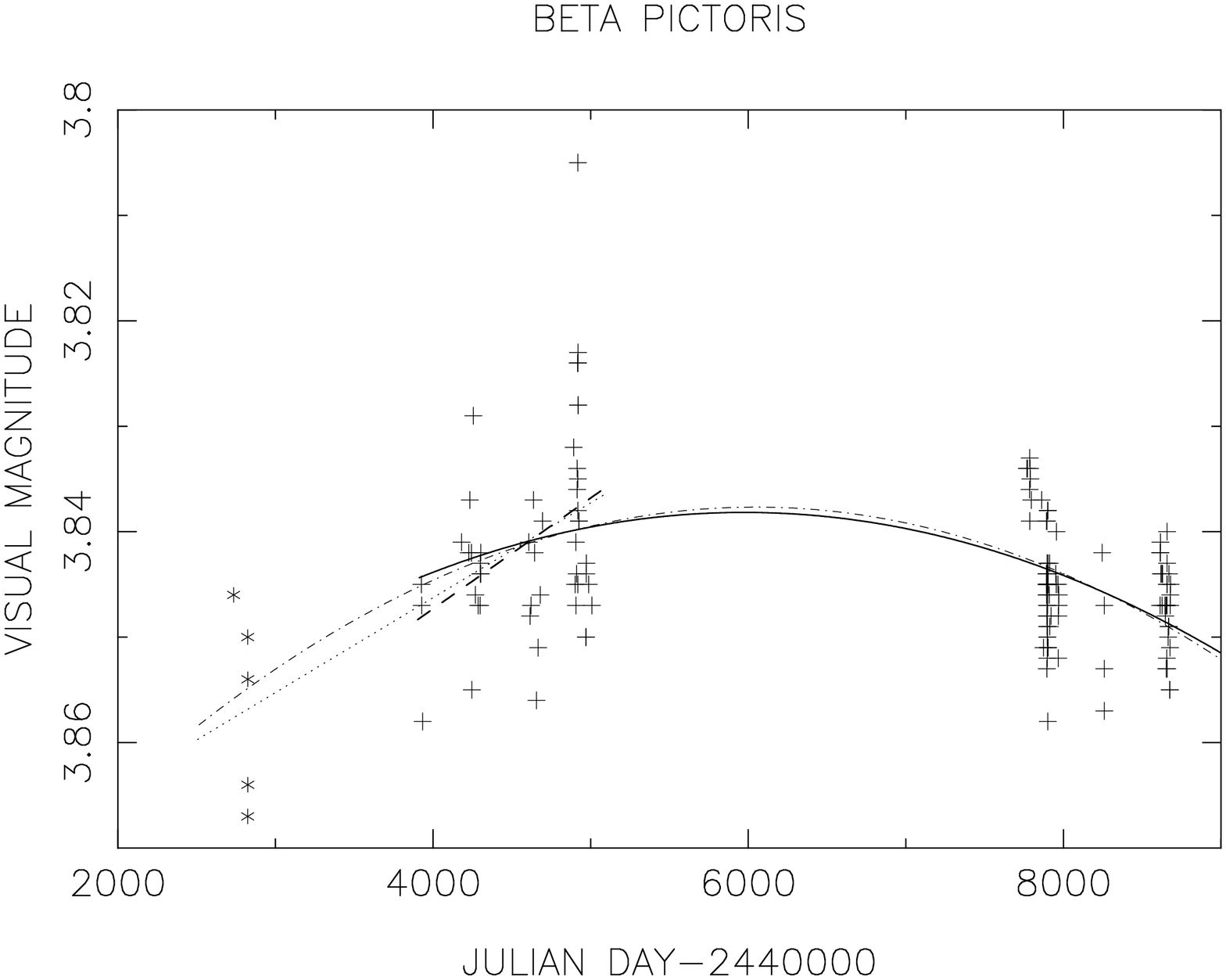,height=\columnwidth,rheight=6truecm,angle=-90}
\else\picplace{6truecm}\setfigdefault\fi
\caption[]{Plot of the magnitudes (crosses) as a function
of Julian day with quality factor greater than or equal to 3,
and curves fit to the data: full line from JD 3925 to JD 8260.
Dot-dash line from JD 2734 to JD 8260.
Dashed line from JD 3925 to JD 5017.
Dotted line from JD 2734 to JD5017.
The data before JD 3000 are represented by asterisk.
With no threshold in quality factor, the fits to the data would be
almost identical}
\label{fit1}
\end{figure}

{}From 1975 to 1982 we have a nearly continuous set of data. But,
in the time span between Nov 18, 1975 to Feb 16, 1976, we have only five
points which are obviously at higher magnitude (Fig.~\ref{fit1}).
These data are very early measurements and are followed by a lack of
measurements during four years.
In order to avoid any bias in the data before JD 3000,
we decided to focus on the more reliable data
from 1979 to 1982

Indeed from JD 3925 to JD 5017, we have 50 measurements of \bp.
During this period of time,
none of the eight other stars analyzed shows significant variations in any
passbands. If we take the 50 first measurements of these eight other stars
after the JD 3900, only for one star (HD 41~692) they show a faint signal
($b_1/s_{b_{1}} = 2.0$) with a chance of 6 \% that $b_1$ is zero. But, since we
consider eight stars, this result can take place 39 times in a series of 100
experiments with eight stars with no variation at all. Thus, we can conclude
that as far as the eight other stars are concerned there are no significant
variations.

However, this is not the case for \bp. It shows a linear variation in the V
magnitude measurements with a slope $b_1=-1.1 \cdot 10^{-5} \ \pm \ 0.4 \cdot
10^{-5}$ mag per day
during about one thousand days. The fitted
curves are plotted in Fig~\ref{fit1}. With $b_1/s_{b_{1}} = 2.7$, the chance
that this result is only due to randomly distributed variations is 0.8 \%.
There is no significant variation in
the color data, this slope occurred with almost the same amplitude in all
wavelengths from 3464~\AA~to 5807~\AA~(U to G filters). The addition of the
$t^2$ term does not improve the fit, since $F=0.01$; the probability that
the measurements depend on the $t^2$ term is less than 9\%.

We shall now mention that the five magnitude measurements before JD
3000 are located exactly over the extrapolation of the variations which
we have described in this section and will further be discussed in
section~\ref{global}. We have plotted the fits with and without these early
measurements. Almost identical fits are found.

\subsection{From 1989 to 1992.}

\label{from 7000}

The fit of this set of data by the first or second degree polynomials gives
that the probability that the slope and the second degree coefficient in the
parabola is zero, is $3\cdot 10^{-3}$ in the first case, and $6\cdot 10^{-6}$
in the second one. We can conclude that we are dealing with strongly
significant signal. However as could be seen from Fig.~\ref{fit1}, this result
is strongly correlated with the first group of data from JD 7767 to 7795
(September 1989) whose magnitudes are smaller than the following ones by about
0$^{\scriptscriptstyle m}\!\!.$01.
In order to check if this effect is really due to the first group of
data, we isolated the data from JD 7873 to the end. In the case where the fit
were made on the whole set of data from JD 7873 to JD 8679, we found almost no
variation: the chance that the slope is zero can then be evaluated to be 18\%.

Moreover, we emphasize that the feature of September 1989 is also present in
the V-B, V$_1$-B and G-B data, with a drop of the V-B color of about
0$^{\scriptscriptstyle m}\!\!.$005,
while the U-B, B$_1$-B and B$_2$-B colors do not show significant
variations. If we consider the first group of data relative to the ones after
JD 7873 (December 1989), the enhancement of the brightness in the short
wavelength seems to be only half that at longer wavelengths. This result is
strongly confirmed by observations of the reference stars, which during the
same
period showed no significant variation.

\subsection{Global analysis of the data.}
\label{global}

Although there is a large discontinuity in the collected data set, we analyzed
them in the same way from a global point of view. With the linear model, we
find a slope with a probability of $10^{-4}$ that it is zero. Furthermore, in
the second model, the probability that the second degree polynomial coefficient
is zero, is $7\cdot 10^{-4}$. This result is presented in Fig.~\ref{fit1} where
we have plotted this parabola with and without the data before 1976. We are
aware that this variation should not be polynomial, but, we do not have enough
information to conclude on the shape of the curve; moreover, the relation
between the first part of the measurements and the second part is not
completely obvious. However, this allows to definitely show that, in the long
term, there is clearly a variation of the magnitude of \bp.

We can check in a very simple manner that the probabilities are well
determined. If we look at the probability that there is a variation (for
example that the slope is different from zero) in a reference star supposed to
be constant, this probability must be uniformly distributed between 0. and 1.
For a group of reference stars, we can therefore sort their probabilities into
ascending order and plot them in this order; then, they must follow a straight
line from 0. to 1.

\begin{figure}[tbp]
\ifthisfig
\psfig{file=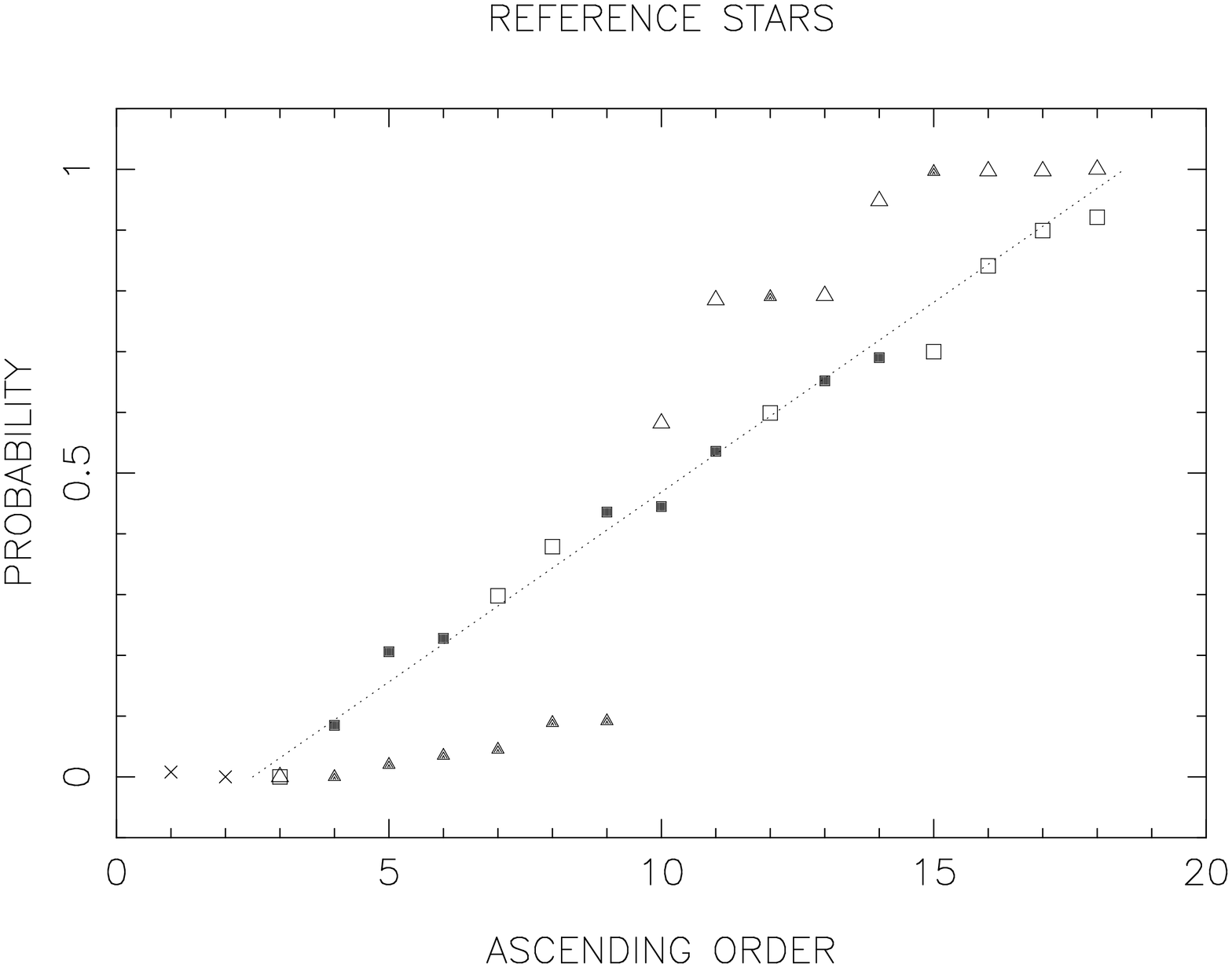,height=\columnwidth,rheight=6truecm,angle=-90}
\else\picplace{6truecm}\setfigdefault\fi
\caption[]{Distribution
of the probabilities that $b_1$ (empty square) and $c_2$
(filled square) are zero among the references stars with a threshold
in the quality factor. It is obvious that these probabilities are uniformly
distributed between 0 and 1.
The triangles represent the same thing with no
threshold in the quality factor. In this case, the probabilities are not
well determined. The two first x are those of \bp\ for which $b_1$ and
$c_2$ are clearly different from zero}
\label{proba}
\end{figure}

The variations were investigated with both models. The result is illustrated on
Fig~\ref{proba}, where we have plotted the probabilities that $b_1$ and $c_2$
are zero in the analysis of the eight reference stars. We see that they
effectively follow the straight line; they are uniformly distributed between 0.
and 1., proving that the magnitudes are purely randomly distributed. Systematic
errors must exist in the data with no threshold on the quality factor since, as
shown in Fig.~\ref{proba}, the probabilities are not as well behaved.

\subsection {IUE Fine Error Sensor observations.}

The Fine Error Sensor (FES) is used by the IUE satellite for target acquisition
and offset guiding (see for example Imhoff, 1989). The FES counts value can be
used to estimate an equivalent V magnitude, using an appropriate calibration
procedure taking into account the stellar color (B-V), the focus step, time
correction and changes of the zero point (Perez \& Loomis 1991).

{}From 1984 until the present, more than two hundred FES counts values have
been
recorded for \bp.
We thus decided to check if this large set of data can be used to
compensate for the lack of measurements. However, since February 1991, the
field of view of the FES is contaminated by a background scattered light
(Monier, 1992). Moreover, from November 1992, this scattered light has strongly
increased with time (Rodriguez Pascual, 1993; Garcia-Lario et al., 1993). We
have thus excluded observations recorded after this date.

In this analysis, we used HD 34~816 ($\lambda Lep$) as an IUE standard star
(see e.g. Huber \& Perez, 1991). For this standard star, we get 91
measurements obtained from 1979 to 1991. The FES derived magnitudes of this
star as a function of time shows absolutely no variation at all, and yields a
mean value of V(FES)=+4.21 with a standard deviation of
0$^{\scriptscriptstyle m}\!\!.$05 which shows the
poor accuracy of these data compared with the Geneva Observatory photometry.

For \bp\, after the different FES corrections, we obtain a set of 201
measurements
from 1984 to 1992. Only 9 measurements has been left
out, they are obviously
wild points out of the distribution. The variance of the remaining 192
measurements of \bp\ (0.0025 mag$^2$) is too large to confirm the features
observed in the Geneva Observatory measurements. We simply note that the points
after JD 8500 (those of 1992) seem to indicate a slow drop in the magnitude of
about 0$^{\scriptscriptstyle m}\!\!.$02; but we are know that since 1991, the
FES data are more and more contaminated by scattered light and the reference
star was not observed during this period. These data only show that a
long term variation of more than 0$^{\scriptscriptstyle m}\!\!.$05 is not
present from 1984 to 1988. Analysis of better quality
measurements is the only way to extract information for this period.

\subsection{Conclusion}

In the long term, we have found that there are two noteworthy
variations. The first one during one thousand days (from JD 3927 to 5017), when
there was an enhancement of the \bp\ brightness of
0$^{\scriptscriptstyle m}\!\!.$011 $\pm$ 0$^{\scriptscriptstyle m}\!\!.$004
from
1979 to 1982, with no associated variations in color. These variations have a
significance level of more than 99.2~\%.
The second variation happened around JD 7780 (September 1989). During this
period the V magnitude of \bp\ was
0$^{\scriptscriptstyle m}\!\!.$01 below the measurements after JD 7767
(December 1989). At shorter wavelengths (U band), the enhancement
was only of 0$^{\scriptscriptstyle m}\!\!.$005.
Finally, the global analysis of the data enable us to definitely show that
there
is clearly a variation of \bp\ brightness on a long time scale.

\section{Short term variations}
\label{short term}

\subsection{Estimation of the variance of the data}

For short term variations, the analysis method used in section~\ref{long term}
could not be applied, because here the variations are present only over a small
part of the data. In order to clear up the possibility of variations on short
time scales, we propose to use a statistical method based on the variance
estimate of different data samples.

In order to compare the variance of the data of different stars, we used the
Snedecor law, which describes the connection between two estimates of the
variance. Among the eight stars we analyzed, we found that there are three
different groups of similar variances. Four stars belong to the first one with
$\sigma^2=1.8\cdot 10^{-5}$ mag$^{2}$ : HD 35~580, HD 41~692, HD 155~450, and
HD 180~885. The second group include two stars with $\sigma^2=3.2\cdot 10^{-5}$
mag$^{2}$. Finally, we have \bp\ with $\sigma^2=6.8\cdot 10^{-5}$, and HD~256,
from JD 7889 to 7903 (29 measurements) where the variance is $\sigma^2=1.3\cdot
10^{-4}$, and from JD 8161 to 8180 (35 measurements) where the variance is
$\sigma^2=1.2\cdot 10^{-5}$. It seems that during the first period of
observation, there was a significant short time variation. This fact will be
discussed in the next section.

As regards \bp, the high value of the variance is due to the data from
Oct 14, 1981 to Feb 17, 1982  (JD 4891 - 5017). In order to eliminate the
effect of long term variations as observed in section~\ref{from 3900}, we
grouped together the measurements taken in a given year, and if we do not take
into account the data of 1981-1982, then we can estimate that the inter-class
variance is $\sigma^2=3.0\cdot 10^{-5}$. This variance is probably enhanced by
the presence of the feature described in sect~\ref{from 7000}, and must thus be
regarded as an upper limit. However we shall use this value in the following
section since an upper limit is what we need.

\subsection{What was going on around \bp\ on Nov 10, 1981?}

Following the results of the previous section, we now suppose that the variance
of the data is $\sigma^2_0=3.0\cdot 10^{-5}$, which is an upper limit. For each
year of observation of each star, we have computed $s^2=\sum (V_i-\bar
V)^{2}/\nu$, where $\nu=n-1$ if $n$ is the number of measurements. The presence
of a signal, with a confidence level of 99\%, is established when
$s^2>\sigma^2_0\chi^2_{\nu}(0.99)/\nu$.

\begin{figure}[tbp]
\ifthisfig
\psfig{file=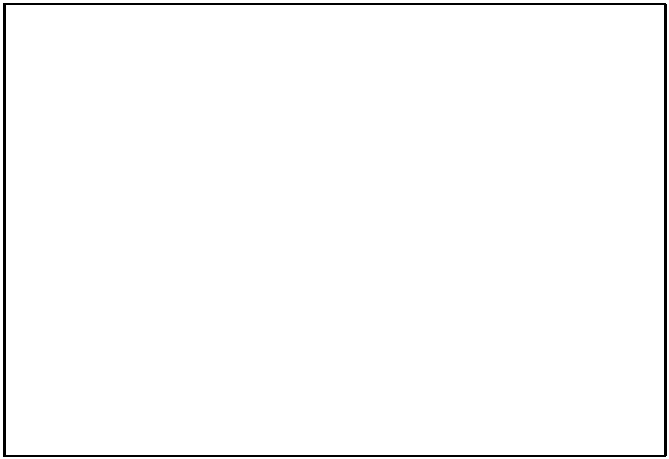,height=\columnwidth,rheight=6truecm,angle=-90}
\else\picplace{6truecm}\setfigdefault\fi
\caption[]{Data of HR~256 (crosses) and their maximum entropy reconstructions
averaged per day (full line). This gives a lower limit of the amplitude of
the variation, assuming that the reconstruction is compatible with the
observations at a confidence level of 99\%}
\label{entro 256}
\end{figure}

\begin{figure}[tbp]
\ifthisfig
\psfig{file=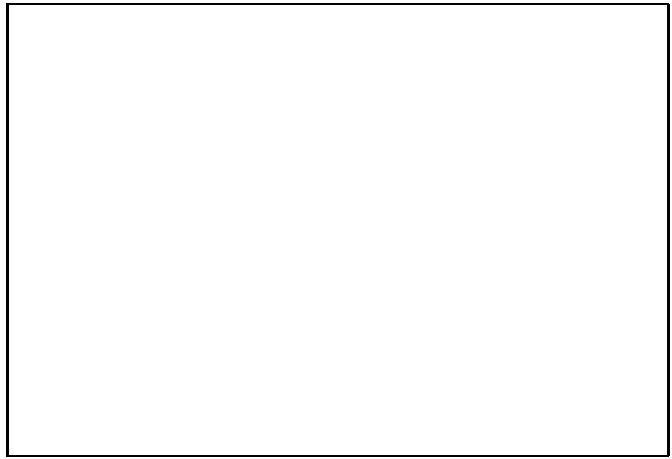,height=\columnwidth,rheight=6truecm,angle=-90}
\else\picplace{6truecm}\setfigdefault\fi
\caption[]{The same as Fig~5.
with the data of \bp\. The most strange data of
Nov 10, 1981 (JD 4918) are represented by circled crosses.
The dashed line represents the mean magnitude of \bp. This reconstruction
gives a lower limit of 0.01 magnitude of variation with a 99\% confidence
level}
\label{entro bp}
\end{figure}

Among all the periods of observation of all the stars (with 45
groups of data), only two groups effectively show a signal with a confidence
level of 99\%: HD~256 from JD 7889 to JD 7903 and \bp\ from JD 4891 to JD 5017.
These data are plotted in Fig~\ref{entro 256} and Fig~\ref{entro bp}. The
chance that the variance is $3.0\cdot 10^{-5}$ could be estimated to be
$1.8\cdot 10^{-14}$ in the first case, and $2.3\cdot 10^{-10}$ in the second
one. For \bp, if we do not take into account the most strange data of JD 4918,
then the chance is still $1.5\cdot 10^{-5}$.

In conclusion, photometric measurements support the presence of a short time
scale variation, with a good level of confidence. We have thus tried to
evaluate the strength of these variations. The estimate of a lower limit of
the amplitude of the variation has been carried out using a maximum entropy
reconstruction of the data, the constraint being that the reconstruction is
compatible with the observations at a confidence level of 99\% : (Skilling \&
Bryan 1984). If $f_i$ is the reconstruction of the data $V_i$ and $A$ the mean
magnitude, then we maximize :
$$
S(f)=-\sum_{j=1}^{N}f_j[\log (f_j/A)-1]
$$
under the constraint that $\sum(f_j-V_j)^{2}/\sigma^2_0 \leq \chi^2_{N}(0.99)$.
The results for HD~256 and \bp\ are illustrated on figures~\ref{entro 256} and
\ref{entro bp}. For \bp, the reconstruction gives a lower limit of 0.01
magnitude of variation. It is extremely interesting to see that the variations
of \bp\ are regular (except for JD 4918), symmetrical and centered on this
peculiar date. We lay stress the fact that, during this particular night, the
atmospheric conditions were very good, and for all the stars observed during
this night the measurements were totally normal.

\section{Discussion}
\label{interpretations}

We now have to deal with the variations with different time scales, for
which we
must explain the durations as well as the amplitudes. We propose some
explanations, but we are aware that they are only tentative in this new field
of study of \bp.

Concerning HD~256 (HR~10), the observed variations
($\Delta V\sim 0^{\scriptscriptstyle m}\!\!.02$)
are not totally surprising since this star shows strong spectroscopic
variations which present some similarity with those of \bp\ (Lagrange-Henri et
al. 1990a, 1990b). However no circumstellar disk was detected (G. Perrin,
private communication). The interpretation of this photometric phenomenon
would be too much speculative,
since photometric variability may be also seen in shell stars.

For \bp, the situation is much better since a number of studies have been made
during the last 10 years. We are aware that we cannot completely exclude the
presence of variations of the intrinsic stellar brightness. However, we think
that these variations are also compatible with disk inhomogeneities for which
optical observations give some indications. Indeed, basing their conclusions on
a midsize grain model, Artymowicz et al. (1989) concluded that the optical
depth calculated from \bp\ outwards to infinity in the midplane of the disk is
$0^{\scriptscriptstyle m}\!\!.03\le\tau_{\Vert}\le
0^{\scriptscriptstyle m}\!\!.07$. This optical depth would be larger in
a model with large grains. A tilt of the disk by less than $5^{o}$ makes that
the extinction of the \bp\ light towards the Earth decreases to only 0.95 of
its
maximum value. Therefore, an inhomogeneity of 10\% to 20\% in the disk
azimuthal distribution of dust can easily produce the variations we have
observed.

Moreover, from a theoretical point of view, it has been demonstrated (Scholl et
al. 1993, Sicardy et al. 1993, Roques et al. 1994) that the presence of a
planet in the \bp\ disk can produce inhomogeneities, such as arcs or
an accumulation of matter following the planet trajectory. These structures can
explain the variations on a long time scale. With a period of more than 2~000
days, the hypothetic planet responsible for such variations must be at more
than 6~AU from the star.

For the short time scale variations, we know that close encounters of particles
with a planet lead to accretion or ejection of these particles into eccentric
orbits (Roques et al. 1994). Thus, the sphere of influence of the planet must
be relatively clear of dust. We propose that the enhancement of brightness of
\bp\ around day 4918 could be due to the passage of this cleared out zone
in front of \bp.
To cover a distance equal to its Hill radius, the planet needs a time $t_H$:
$$
t_H=32\left(\frac{M_p}{M_{J}}\right)^{1/3}
      \left(\frac{D}{5 {\rm AU}}\right)^{3/2} \ {\rm days}
$$
Here $M_p$ is the mass of the planet, $M_J$ is the mass of Jupiter, and $D$ is
the distance from the planet to \bp. Note that this time is similar to the
duration of the phenomenon around the day 4918. Moreover, if we assume
$R_{\beta \ Pic}=1.2R_{\odot}$ (Paresce 1991), an occultation by a planet
crossing a diameter of \bp\ will last a time $t_{\star}$, where
$t_{\star}\le 1.1 (D/5 {\rm AU})^{1/2}\ {\rm days}$, and will
decrease the
luminosity of the star by $\delta V = 8\cdot 10^{-3} (R_p/R_J)^{2}$. The
duration of the JD 4918 phenomenon constrains $D$ to be greater than 0.08~AU.
No upper limit can be given. For a planet with a radius 1.1 times the radius of
Jupiter, we find $\delta V= 0^{\scriptscriptstyle m}\!\!.01$, and
$\delta V=0^{\scriptscriptstyle m}\!\!.02$ with a planet 1.6 times
bigger than Jupiter. Thus, the measurements of JD 4918 could be explained by an
occultation of the star by a planet or a group of planets (planetesimals?)
which cover a little more than an area similar to the surface of Jupiter.

\section{Conclusion}

{}From the analysis of \bp\ photometry measurements collected by the Geneva
Observatory from 1975 to 1992, we have shown evidence of variations of the
apparent magnitude of the star, on long as well as on short time scales.

On long time scales, we proved that the brightness of \bp\ decreased by
0$^{\scriptscriptstyle m}\!\!.$011
$\pm$ 0$^{\scriptscriptstyle m}\!\!.$004
over one thousand days. Unfortunately, the quality of the IUE
measurements is too low to compensate for the lack of Geneva Observatory
measurements between 1983 and 1988. However, when we consider all the
measurements, the chance that there is no variation at all can be estimated to
be less than $10^{-4}$. This result is strongly confirmed by the comparison
with the other stars on which the same tests were performed and which showed no
long term variations at all.

On short time scales, peculiar variations are observed during about 30 days.
These variations are regular, symmetrical and centered on a very particular day
(JD 4918) on which variations may be reach 0$^{\scriptscriptstyle m}\!\!.$04.

Except for the variations around JD 7780, they do not show color effects. The
variations have the same amplitude in all wavelengths from 3464~\AA\ to
5807~\AA\ (U to G filters), indicating that probably large particles are
responsible for these variations.

We are aware that the explanations proposed here are somewhat bold, all the
more so as we do not have a lot of information. For the results presented here,
it
is not possible {\it a priori} to exclude a stellar origin for the observed
photometric variations, rather than the proposal of the variations arising from
local density variations of the disk. However, it is also difficult to
show this. Thus, we have looked at an other hypothesis based on
observational and theoretical information on the \bp\ disk. These variations,
on long as well as on short time scales, could be due to the presence of
inhomogeneities in the disk which could be related to the presence of planets
or planetesimals. It seems to be very important to continue such observations
to confirm the existence of the variations, and, with the help of other
observational techniques (spectroscopy, high resolution observations) build a
unified model for the fascinating environment of this star.

\begin{acknowledgements}

We would like to express our gratitude to Dr. H.Scholl for very fruitful
discussions.
We are particularly indebted to Dr. E.Guinan for improving the presentation
and discussion on the results.
Our thanks go also to Dr. J.Schneider for his comments and suggestions and
Dr. M. Friedjung for improving the manuscript.

\end{acknowledgements}

\end{document}